# 1/F NOISE IN SPATIALLY EXTENDED SYSTEMS WITH ORDER-DISORDER PHASE TRANSITIONS

K.Staliunas
Physikalisch Technische Bundesanstalt, 38116 Braunschweig, Germany
tel.: +49-531-5924482, Fax: +49-531-5924423, E-mail: Kestutis.Staliunas@PTB.DE

**Abstract**

Noise power spectra in spatially extended dynamical systems are investigated, using as a model the Complex Ginzburg-Landau equation with a stochastic term. Analytical and numerical investigations show that the temporal noise spectra are of $1/f^\alpha$ form, where $\alpha = 2 - D/2$ with $D$ the spatial dimension of the system. This suggests that nonequilibrium order-disorder phase transitions may play a role for the universally observed $1/f$ noise.

Order - disorder phase transitions in spatially extended nonlinear systems subjected to additive noise, are universally described by complex Ginzburg - Landau equation (CGLE):

$$\frac{\partial A}{\partial t} = pA - (1+ic)|A|^2 A + (1+ib)\nabla^2 A + \Gamma(\mathbf{r},t) \qquad (1)$$

where $A(\mathbf{r},t)$ is the order parameter defined in n-dimensional space $\mathbf{r}$, and evolving with time $t$. $p$ is the control parameter (order - disorder transition occurs at $p = 0$). The Laplacian $\nabla^2 A$ represents nonlocality in the system, and $\Gamma(\mathbf{r},t)$ is the noise, $\delta$ - correlated in space and time of power (temperature) $T$: $\langle \Gamma(\mathbf{r}_1,t_1) \cdot \Gamma^*(\mathbf{r}_2,t_2) \rangle = 2T \cdot \delta(\mathbf{r}_1 - \mathbf{r}_2)\delta(t_1 - t_2)$

Below the transition threshold ($p < 0$) CGLE (1) yields a disordered state: the complex-valued order parameter $A(\mathbf{r},t)$ is essentially a noise filtered in space and time, with exponential (thermal) intensity distribution. Above the transition threshold ($p > 0$) (1) yields an ordered, or coherent state (or a condensate) in modulationally stable case, with the order parameter distributed near its mean value $\langle |A|^2 \rangle = p$.

CGLE (1), with complex-valued coefficients, has been derived systematically for many systems describing second order phase transitions in the presence of noise, e.g.: for lasers with spatial degrees of freedom [1], where $A(\mathbf{r},t)$ is proportional to the amplitude of the optical field, and $\Gamma(\mathbf{r},t)$ corresponds to the vacuum or thermal fluctuations; for finite temperature superfluids [2], and for finite temperature Bose-Einstein condensates [3], where $A(\mathbf{r},t)$ is the wave-function of the condensate, and $\Gamma(\mathbf{r},t)$ corresponds to the fluctuations of thermal bath. The CGLE with real-valued coefficients $b = c = 0$, has been also systematically derived as the amplitude equation for stripe patterns in nonequilibrium dynamical systems [4], where the amplitude and phase of the order parameter corresponds to the amplitude- and phase modulations of the roll patterns respectively.

The CGLE, with real-valued coefficients $b = c = 0$, can be written phenomenologically as the normal form, or the minimal equation describing universally nonequilibrium order - disorder phase transitions in the presence of noise [5]: the first two terms $(pA - |A|^2 A)$ approximate in the lowest order a supercritical Hopf bifurcation, and the diffusion term $\nabla^2 A$ describes the simplest possible nonlocality. The complex-valued character of the order parameter $A(\mathbf{r},t)$ is important: every ordered, or coherent state, both in classical or quantum mechanics, is characterized not only by the modulus of the order parameter, bus also by its phase. Then, as the real Ginzburg-Landau equation is the normal form of second order phase transitions between two arbitrary states [6], the CGLE (1) can serve the as normal form of the second order phase transitions between ordered and disordered states of the matter.

It is shown in this letter, that the noise spectra of the CGLE with the real-valued coefficients $b = c = 0$ are of $1/f^\alpha$ form, where the value of $\alpha$ depends only on the dimension of the system. As the CGLE is a universal equation, describing not necessarily a concrete system, but rather generally order-disorder phase transitions, then this may explain the omnipresence of $1/f$ noise in nature. $1/f$ noise has been observed in many very different types of systems, from physics, technology, biology, astrophysics, geophysics and sociology [7]. There exist physical models of $1/f$ noise for some concrete physical systems but those are usually specialized, and can thus not explain the omnipresence of $1/f$ noise. One exception is perhaps the self-organized criticality [8] where $1/f$ noise appears universally for a large class of critical systems. In the present letter the omnipresence of $1/f$ noise is explained not only for critical systems, as in [8], but for arbitrary supercritical spatially extended ordered dynamical systems. In other words - the $1/f$ noise is predicted here to occur where ever an ordered or coherent state is subjected to additive noise.

For analytical treatment it is assumed, that the system is sufficiently far away from the order - disorder phase transition: $p \gg T$. Then the homogeneous component $|A_0| = \sqrt{p}$ is dominating, and one can look for a solution of (1) in the form of a perturbed homogeneous state: $A(\mathbf{r},t) = A_0 + a(\mathbf{r},t)$. Linearizing (1) around the stationary homogeneous solution one obtains:

$$\frac{\partial a}{\partial t} = -p(a + a^*) + \nabla^2 a + \Gamma(\mathbf{r},t) \qquad (2)$$

and its complex conjugate. Diagonalisation of (2) is straightforward: $b_+ = (a + a^*)/\sqrt{2}$ and $b_- = (a - a^*)/\sqrt{2}$, which yields:

$$\frac{\partial b_+}{\partial t} = -2pb_+ + \nabla^2 b_+ + \Gamma_+(\mathbf{r},t) \qquad (3.\text{a})$$

$$\frac{\partial b_-}{\partial t} = \nabla^2 b_- + \Gamma_-(\mathbf{r},t) \qquad (3.\text{b})$$

This shows, that:

a) all amplitude fluctuations $b_+$ decay above the phase-transition point with a decay rate: $\lambda_+ = -2p - k^2$, where $k$ is the spatial wavenumber of the perturbation mode. Asymptotically long-lived amplitude perturbations are possible only at the phase transition point (in a critical state), but never above it;

b) the phase fluctuations $b_-$ decay with a rate $\lambda_- = -k^2$, which means that the long-wavelength modes decay asymptotically slowly, with a decay rate approaching zero for $k \to 0$.

From (3) one can calculate spatio-temporal noise spectra, by rewriting (3) in terms of the spatial and temporal Fourier components $b_\pm(\mathbf{r},t) = \int b_\pm(\mathbf{k},\omega) d\omega d\mathbf{k}$: This leads to:

$$b_+(\mathbf{k},\omega) = \frac{\Gamma_+(\mathbf{k},\omega)}{i\omega + \mathbf{k}^2 + 2p} \qquad (4.\text{a})$$

$$b_-(\mathbf{k},\omega) = \frac{\Gamma_-(\mathbf{k},\omega)}{i\omega + \mathbf{k}^2} \qquad (4.\text{b})$$

The spatio-temporal power spectra are:

$$S_+(\mathbf{k},\omega) = |b_+(\mathbf{k},\omega)|^2 = \frac{|\Gamma_+(\mathbf{k},\omega)|^2}{\omega^2 + (2p + \mathbf{k}^2)^2} \qquad (5.a)$$

$$S_-(\mathbf{k},\omega) = |b_-(\mathbf{k},\omega)|^2 = \frac{|\Gamma_-(\mathbf{k},\omega)|^2}{\omega^2 + \mathbf{k}^4} \qquad (5.b)$$

for the amplitude and phase fluctuations correspondingly. Assuming the $\delta$ - correlated noise in space and time, $|\Gamma_\pm(\mathbf{k},\omega)|^2$ are simply proportional to the temperature $T$ of the random force.

The temporal spectrum is obtained by integration (5) over all possible wavevectors $\mathbf{k}$: $S_{tot}(\omega) = S_{amplitude}(\omega) + S_{phase}(\omega) = \int S_+(\mathbf{k},\omega)d\mathbf{k} + \int S_-(\mathbf{k},\omega)d\mathbf{k}$. (The total power spectrum here is the sum of amplitude and phase power spectra, since the spectral components $b_\pm(\mathbf{r},t)$ are mutually uncorrelated, as follows from (4)). The integration is performed for clarity separately for amplitude and phase fluctuations. In the case of one spatial dimension the integration yields:

$$S_{+1D}(\omega) = \int_{-\infty}^{\infty} \frac{T}{\omega^2 + (2p+k^2)^2} dk = \frac{T\pi}{\omega} \text{Im}\left[(2p-i\omega)^{-1/2}\right] \qquad (6.a)$$

$$S_{-1D}(\omega) = \int_{-\infty}^{\infty} \frac{T}{\omega^2 + k^4} dk = \frac{T\pi}{2^{1/2}\omega^{3/2}}. \qquad (6.b)$$

This means that the spectrum of phase fluctuations is precisely of the form $1/\omega^{3/2}$ in all frequency ranges (6.b). The spectrum of amplitude fluctuations is more complicated: in the limit of large frequencies $|\omega| \gg 2p$ it is equal to the phase spectrum $S_{+1D}(\omega) = S_{-1D}(\omega)$, as follows from (6.a). In the limit of small frequencies $|\omega| \ll 2p$, the amplitude power spectrum saturates to: $S_{+1D}(\omega \approx 0) = T\pi/(2\cdot(2p)^{3/2})$. In this way (6.a) yields a Lorenz-like spectrum of amplitude fluctuations, which however does not follow the usual $1/f^2$ law, but the $1/f^{3/2}$ law for the systems extended in one-dimensional space for asymptotically large frequencies.

The integration of (5) in two spatial dimensions yields:

$$S_{+2D}(\omega) = 2\pi \int_0^{\infty} \frac{T}{\omega^2 + (2p+k^2)^2} k\,dk = \frac{T\pi}{2\omega}(\pi - 2\cdot arctg(2p/\omega)) \qquad (7.a)$$

$$S_{-2D}(\omega) = 2\pi \int_0^{\infty} \frac{T}{\omega^2 + k^4} k\,dk = \frac{T\pi^2}{2\omega}. \qquad (7.b)$$

This yields a power spectrum of phase fluctuations in the form $1/\omega$ in all frequency ranges, and a Lorenz-like spectrum of amplitude fluctuations decaying with $\alpha = 1$ law at high frequencies. It saturates to $S_{+2D}(\omega \approx 0) = T\pi/(2p)$ at low frequencies.

Finally, the integration of (5) in 3D yields:

$$S_{+3D}(\omega) = 4\pi \int_0^{\infty} \frac{T}{\omega^2 + (2p+k^2)^2} k^2\,dk = \frac{2T\pi^2}{\omega} \text{Im}\left[(2p+i\omega)^{1/2}\right] \qquad (8.a)$$

$$S_{-3D}(\omega) = 4\pi \int_0^{\infty} \frac{T}{\omega^2 + k^4} k^2\,dk = \frac{T\pi^2 2^{1/2}}{\omega^{1/2}}. \qquad (8.b)$$

This results in spectrum of the form $1/\omega^{1/2}$ for the phase fluctuations and a Lorenz-like spectrum of intensity fluctuations with $\alpha = 1/2$ at high frequencies, and saturating to $S_{+3D}(\omega \approx 0) = T\pi^2/(2p)^{1/2}$ at low frequencies.

Generalizing, the power spectra of phase fluctuations for D-dimensional systems are $S_{-nD}(f) \approx 1/f^{\alpha}$ with $\alpha = 2 - D/2$. For the amplitude fluctuations one obtains Lorenz-like power spectra, with the usual saturation for small frequencies, and with $1/f^{\alpha}$ decay for high frequencies with $\alpha = 2 - D/2$ depending on the dimension of the system.

The spectral densities calculated above (6)-(8) were compared to those obtained by numerical integration of CGLE (1) in 1, 2, and 3 spatial dimensions. The total spectra were numerically calculated, and were plotted in Fig.1.a. The $1/f^{\alpha}$ character of noise spectrum is most clearly seen in the case of a 1D system (here $\alpha = 3/2$). In two dimensions the $1/f^{\alpha}$ noise ($\alpha = 1$) is visible over almost three decades of frequency, and in three dimensions ($\alpha = 1/2$) over two decades. The dashed lines in Fig.1.a indicate the corresponding slopes.

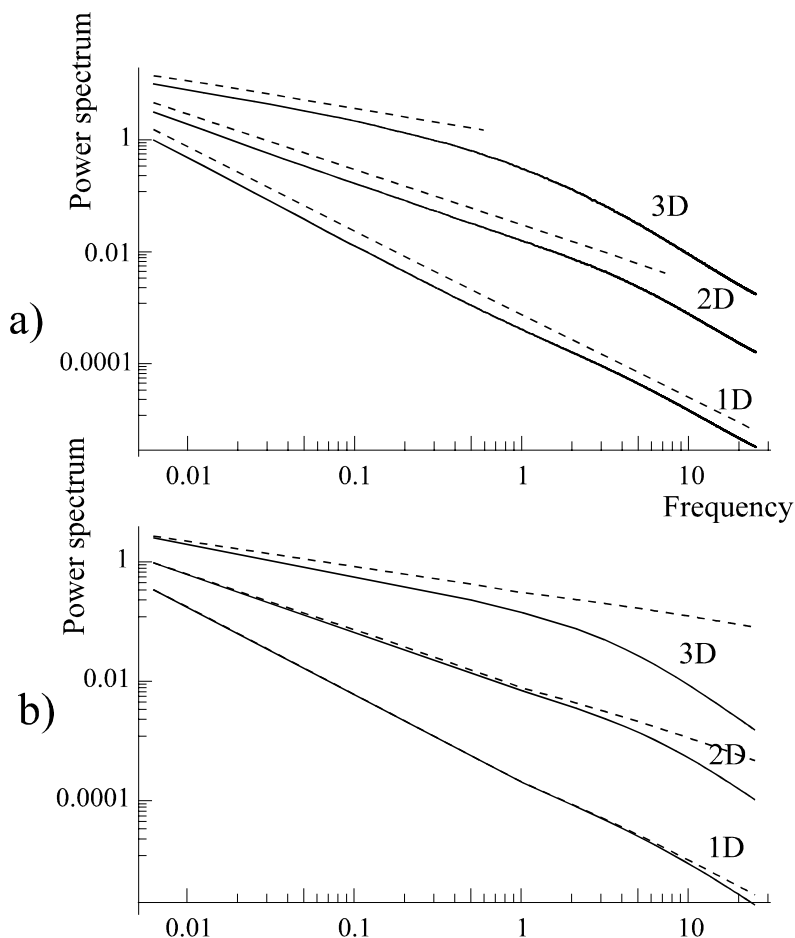

Fig.1.
a) Total noise power spectra in 1, 2 and 3 spatial dimensions, as obtained by numerical integration of CGLE (1) with real-valued coefficients $b = c = 0$, and $p = 1$. Dashed lines show the slopes $\alpha = 1/2$, $\alpha = 1$, and $\alpha = 3/2$. The spectra are arbitrary displaced vertically to distinguish between them.
b) analytically calculated spectra from integration of (5). Dashed curves correspond to integration over the whole space of **k**-vectors, solid curves correspond to integration in the limited space $0 < |\mathbf{k}| < k_{max}$. $k_{max}$ correspond to cases calculated numerically, namely = $2\pi*64$, $2\pi*32$, and $2\pi*16$ for 1D, 2D and 3D systems respectively. Integration period is 1000; the averaging was performed over 2500 realizations. Periodic boundary conditions were used.

The main obstacle to observe numerically the $1/f^{\alpha}$ noise in the entire frequency range is the discretization of the spatial coordinate in the integration scheme. Due to numerical discretization, a limited number of spatial modes is taken into account only, thus the higher spatial modes are truncated. To check the above discussed effects of discretization the integrals (5) were calculated analytically, however, integrating not over the whole space: $0 < |\mathbf{k}| < \infty$ (like performed to obtain the analytical expressions (6)-(8)), but integrating over a limited region of transverse wavenumbers: $0 < |\mathbf{k}| < k_{max}$. The analytical expressions are complicated. In Fig.1.b the corresponding plots are given. The integration taking into account all possible values of **k** lead to $1/f^{\alpha}$ spectra (dashed curves), with hardly visible small discrepancies from the power law at $|\omega| \approx 2p$. Truncating the spatial spectra to the values $k_{max}$ (chosen from the conditions used in the corresponding numerical integration) yields the discrepancies at larger frequencies. Very similar discrepancies are indeed observed in the spectra obtained numerically.

Concluding, the results show that the power spectra of phase fluctuations are of $1/f^{\alpha}$ form in the entire frequency range. The power spectra of amplitude fluctuations are Lorenz-like with $1/f^{\alpha}$ decay for high frequencies. The coefficient $\alpha = 2 - D/2$ depends on the dimension of space.

In Fig.1 the sum of the amplitude and phase noise spectra are plotted. Is this what one usually observes in experiments on spatially extended systems?

It seems that mostly the total power spectra are observed. E.g. in the case of spatially extended lasers the amplitude and phase fluctuations mix in free propagation of the light beam, and what a photodetector measures is the total noise. The thermal noise in the spatially extended electronic devices (e.g. a resistor) stems equally from the amplitude and phase noise: the phase noise in an ordered ensemble of charge carriers results in the spatial fluctuations of the electric charge, thus on the voltage registered. Finally, the intensity fluctuations at a given location of an ordered pattern (e.g. zebra, or hexagonal) result equally from the amplitude and phase fluctuations of the order parameter. The last is illustrated by the analysis of stripe patterns of the Swift-Hohenberg equation:

$$\frac{\partial U}{\partial t} = U - U^3 - (\Delta + \nabla^2)^2 U + \Gamma(\mathbf{r},t) . \qquad (9)$$

$U(\mathbf{r},t)$ is the real-valued order parameter, $\Delta$ is the detuning parameter, and $\Gamma(\mathbf{r},t)$ is the real valued $\delta$- correlated noise. The Swift-Hohenberg equation (9) is the simplest model which yields stripe patterns: $U(\mathbf{r}) \approx \exp[i\mathbf{k}\mathbf{r}] + \exp[-i\mathbf{k}\mathbf{r}]$ with an arbitrarily oriented wavevector with modulus dependent on the detuning: $\mathbf{k} = \sqrt{\Delta}$, for sufficiently large values of the detuning [9]. The amplitude equation for the modulation of the stripes: $U(\mathbf{r},t) = A(\mathbf{r},t)\exp[i\mathbf{k}\mathbf{r}] + c.c.$ of Swift-Hohenberg equation (9), (also of other models supporting stripe patterns) is the CGLE (1) [4]. Here the modulus of the order parameter $A(\mathbf{r},t)$ corresponds to the local stripe amplitude, and the phase of the order parameter to the local spatial shift of the stripes. In this way the amplitude and phase fluctuations of the order parameter $A(\mathbf{r},t)$ participate equally in the locally measured fluctuations of order parameter $U(\mathbf{r},t)$ in the presence of stripes (and for an arbitrary ordered patterns, like hexagons, e.t.c.).

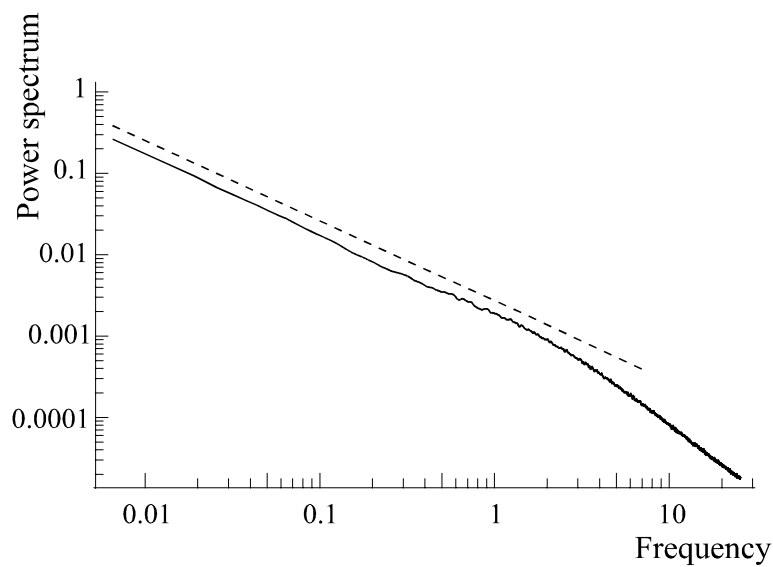

Fig.2. Intensity noise power spectrum in the case of two spatial dimensions, as obtained by numerical integration of Swift-Hohenberg equation (9) at a detuning $\Delta = 4$. The solution is a fluctuating stripe pattern. Integration period is 1000; the averaging was performed over 1500 realizations. Periodic boundary conditions were used.

Fig.2 shows the noise spectrum of the intensity fluctuations of the order parameter $U(\mathbf{r},t)$ as obtained by numerical integration of the Swift-Hohenberg equation (9) in two spatial dimensions. The power spectrum shows clearly the 1/f noise in a large frequency region (as allowed by discretization of the spatial coordinates).

All ordered states (patterns) in the nature are D-dimensional, with $1 < D < 3$. According to calculations above this correspond to the power indices of $1/f^\alpha$ noise $1/2 < \alpha < 3/2$, which correspond well to experimentally observed powers of the 1/f noise in the nature [7].

The above predictions also lead to conclusion, that the ordered state is absolutely stable only in 3D case, where the noise is $1/f^{1/2}$. The integration of the noise spectrum over all frequency range does not lead to infinity in this 3D case, thus the fluctuations of the order parameter are finite sufficiently away from the order-disorder phase transition. In contrary, the integrals of $1/f^\alpha$ noise spectra in 1D and 2D with $\alpha \leq 1$ yield infinities. This means, that the ordered states of the infinitely extended system in less than three dimensions are never absolutely stable: even arbitrary far away from the order-disorder transition threshold there continuously occur fluctuations of the order parameter eventually destroying the ordered state in cases of 1D and 2D.

Extending the above analysis to the case of zero-dimensional system the Lorenz spectrum is easily retrieved, with $\alpha = 2$, which is a well known result for compact integrating systems, and which may serve as some check of the results reported in the paper. The noise spectra in four-dimensional cases diverge weakly (logarithmically) for small frequencies. The noise spectra in five- and higher-dimensional cases are of $f^\alpha$ form, thus not diverge for small frequencies at all.

The fluctuation spectra were calculated for the CGLE with real-valued coefficients $b = c = 0$, however the results can be easily extended to the case of CGLE with complex-valued coefficients in modulational stability range: $1 + bc > 0$. The eigenvalues for linearized equation (analog of (2)) are: $\lambda_+ = -2p - k^2(1-bc)$ and $\lambda_- = -k^2(1+bc)$ for amplitude and phase perturbations, in the long wave limit $k^2 \ll 1$. This generates the linear Langevin equations similar to (3), and eventually leads to the same $1/f^\alpha$ spectra. This extends the omnipresence of $1/f^\alpha$ from predominantly dissipative systems $b, c \ll 1$, such as e.g, reaction-diffusion systems, to predominantly conservative systems $b, c \gg 1$, such as e.g. Bose-Einstein condensates.

The work has been supported by Sonderforschugsbereich 407 of Deutsche Forschungsgemeinschaft. The discusions with C.O.Weiss and A.Berzanskis are acknowledged.